\documentstyle[epsf,rotate,twocolumn,aps,pre]{revtex}

\begin{document}
\draft

\twocolumn[\hsize\textwidth\columnwidth\hsize\csname
@twocolumnfalse\endcsname

\title{
Mode coupling theory for molecular liquids: What can we learn from a
system of hard ellipsoids?}

\author{M. Letz, R.~Schilling\\
Johannes Gutenberg-Universit\"at, 55099 Mainz, Germany} 

\date{\today}

\maketitle

\begin{abstract}
Molecular fluids show rich and complicated dynamics close to the glass
transition. Some of these observations are related to the fact that
translational and orientational degrees of freedom couple in nontrivial
ways. A model system which can serve as a paradigm to understand
these couplings is a system of hard ellipsoids of revolution. To test
this we compare at the ideal glass transition the static molecular correlators of a
linear A-B Lennard--Jones molecule obtained from a molecular dynamics
simulation with a selected fluid of hard
ellipsoids for which the static correlators have been obtained using
Percus--Yevick theory. We also demonstrate that the critical
non-ergodicity parameters obtained from molecular mode coupling theory
for both systems show a remarkable similarity
at the glass transition, provided the aspect ratio is chosen properly. Therefore we
conclude that a system of hard ellipsoids can indeed be used
to understand part of the essential behaviour of such a simple
molecular system like the A-B Lennard-Jones molecules in the vicinity of the
ideal glass transition. 
\end{abstract}
]
\narrowtext

\section{Introduction}
\label{sec:intro}
When a fluid is driven towards a glass
transition either by quenching down the temperature (super-cooling) or
by increasing the density a critical slowing down of the dynamics
occurs without a diverging length scale in the system.\\
However there are only a few systems where the full dynamics of the
glass transition is understood in 
detail and where it is possible to make predictions for the critical
density and 
temperature for the ideal glass transition from a microscopic theory.
Systems where such a transition is well understood are
colloidal suspensions \cite{vanmegen95}. Their liquid state is well
described by a hard sphere system  and the static structure factor
calculated by Percus--Yevick approximation has been used as an input
for the mode coupling theory (MCT) \cite{bengtzelius84} to calculate
the dynamics  
near  the ideal glass transition \cite{fuchs92,fuchs92b}. 
Within idealized MCT this transition takes place at a critical
temperature $T_c$ or a critical density  
$\rho_c$ at which a breaking of ergodicity occurs. 
For a simple glass former like a
colloidal system the mode coupling equations describe the dynamics in
great detail, explain why the transition is dominated by the cage
effect and their solutions are in good agreement with experimental data
\cite{vanmegen95}.\\
Most  glass formers do not have such a simple structure like
hard spheres. A real glass former is usually a
molecular system and shows a wide variety of additional structure in
the dynamics (multiple microscopic peaks, fast and slow $\beta$ peaks,
$\alpha, \alpha'$ peaks) where a detailed understanding is lacking so far. 
A large part of this
additional structure is caused by the coupling of translational and
orientational degrees of freedom. There are different possibilities to
get a deeper insight into such couplings.\\
The first possibility to study the relaxation in the super-cooled regime 
is to model a specific glass former as
accurately as possible using model potentials, and then to perform computer
simulations. This has been done, e.g. for water \cite{fabbian98b,gallo96}, for
silicate glass \cite{kob98} or for a linear
A-B-molecule\cite{kaemmerer97}.\\
A second possibility to get insight  into the statics as well as in the
dynamics of a glass former is to study model systems which are as
simple as possible but keep the essential mechanism -- in our case the
interplay between translational and rotational degrees of freedom. Such a
model system which can be used is a system of hard ellipsoids of
revolution.\\[0.3cm]
The aim of this paper is to answer the question how far a system of
hard ellipsoids can help to understand basic
mechanisms present in molecular glass formers.\\
We therefore compare in detail the static structure of the translational and orientational degrees of freedom obtained from a
molecular dynamics simulation for an A-B-Lennard Jones molecule with
results for hard ellipsoids of revolution. Using the static correlators
as an input into the mode coupling equations we also compare the
results for the critical non--ergodicity parameters with each other.

\section{Mode-coupling equations}

As a first step  towards finally calculating the full dynamics
we want to test for both systems  the occurrence of ergodicity breaking. Therefore
the MCT-- equations are solved in the limit $t \longrightarrow \infty$
The non-ergodicity parameters are given by $\left ( {\bf F}(q,m) \right )_{ll'} \equiv
F_{ll'}(q,m) = \lim_{t \rightarrow \infty} S_{ll'}(q,m;t)$ where
$S_{ll'}(q,m;t)$ is the time dependent tensorial density correlation function.
The indices $l,l'$ and $m$ refer to expansion coefficients of a product of
spherical harmonics $Y_l^m(\Omega)$ and $Y_{l'}^{m'}(\Omega')$. Since
we have used the q-frame representation 
for the correlators, i.e. $\bf q$ has been chosen along the z--axis,
the correlators are diagonal in $m$ and $m'$ and are real
\cite{schilling97}. The reader should also note that 
the existence of a rotational axis of symmetry  of the species of both systems allows to use the spherical harmonics, only.
Using the results from ref. \cite{schilling97} and \cite{theis97,fabbian99} one
finds the molecular MCT--equations for these parameters:
\begin{eqnarray}
\label{eq:mctanf}
\lefteqn{
{\bf F}(q,m) = \nonumber }\\ &&  \left [{\bf S}^{-1}(q,m) + {\bf S}^{-1}(q,m) 
{\bf \cal K}(q,m; \{ {\bf F}(q,m) \} ) {\bf S}^{-1}(q,m) \right ] ^{-1}      
\end{eqnarray}
where the matrix elements of the functional ${\bf \cal K}(q,m; {\bf
F}(q,m))$  are given by:
\begin{eqnarray}
\lefteqn{
{\bf \cal K}(q,m; \{ {\bf F}(q,m) \} ) = \nonumber } \\ &&
\sum_{\alpha=T,R} \sum_{\alpha'=T,R}
q_l^{\alpha}(q) \left (
\left [ {\bf m}(q,m; \{ {\bf F}(q,m) \} ) \right ] ^{-1} \right )
^{\alpha \alpha'} _{ll'} q^{\alpha'}_{l'}(q)
\end{eqnarray}  
with the MCT--polynomial:
\begin{eqnarray}
\label{eq:mctker}
\label{eq:mctend}
\lefteqn{
\left ( {\bf m}(q,m; \{ {\bf F}(q,m) \} )  \right )
^{\alpha \alpha'} _{ll'} \equiv  
{\bf m}^{\alpha \alpha'} _{ll'}(q,m; \{ {\bf F}(q,m) \} )} 
\nonumber \\  &=&  
\frac{1}{2N} \sum_{q_1 q_2 \atop m_1 m_2}
\sum_{l_1 l_2 \atop l_1' l_2'}
V^{\alpha \alpha'}_{ll';l_1l_1',l_2l_2'}(q,m|q_1,m_1;q_2,m_2)
\nonumber \\ &&\mbox{\hspace*{1cm}}
F_{l_1l_1'}(q_1,m_1)
F_{l_2l_2'}(q_2,m_2)
\end{eqnarray}
and
\begin{equation}
q_l^{\alpha}(q) = \left \{ \begin{array}{ccc}
q&,& \alpha = T \\
\sqrt{l(l+1)}&,& \alpha = R 
\end{array}
\right .
\end{equation}
The explicit expressions for the vertices $V^{\alpha,\alpha'}$ for
arbitrary ${\bf q}$ can be found in ref. \cite{schilling97} and for
the q-frame in ref. \cite{theis97,fabbian99}. 
The vertices $V^{\alpha \alpha'}$  only
depend on the static correlators $S_{ll'}(q,m)$ 
and the direct correlation functions $c_{ll'}(q,m)$ which 
are related to each other by the Ornstein--Zernike equation:
\begin{equation}
{\bf S}(q,m) = \left [ {\bf 1} -\frac{\rho}{4 \pi} {\bf c}(q,m) \right
] ^{-1}
\end{equation}
where $\rho$ is the number density.
\section{Comparison}
For two particular systems the above equations have been solved. The
{\it first} one is the diatomic linear A-B molecule of
ref. \cite{kaemmerer97} were the static correlators had been obtained by
a molecular dynamics simulation.
The parameters of the Lennard-Jones potentials 
had been chosen such that the diameter of  atom A and B were equal to
1 and 0.95 (in Lennard-Jones units), respectively. Such an asymmetry
is mainly needed to prevent 
crystallization. The distance between the two centers of the atoms was
0.5 .\\
The {\it second} system  is a fluid of
prolate hard ellipsoids of revolution.
Here the aspect ratio
X$_0=\frac{a}{b}$
which is the ratio between the major axis $a$ and the minor
axis $b$ of the ellipsoids has to be chosen. If one wants to model
the Lennard--Jones molecule
by a hard ellipsoid the choice of the Lennard--Jones radii implies that X$_0= 1.5$.
Note that X$_0$ is not a fitting parameter. We also stress that the A--B molecules do not have a head--tail symmetry, in contrast to the ellipsoids. The value for X$_0$ is in a parameter range where no ordering due to a nematic phase is expected
\cite {frenkel84,letzallen99}.
The static correlators are  obtained using 
Percus--Yevick (PY) theory. Using PY theory automatically prevents
crystallization since it is unable to yield a periodically ordered
solution. PY, however, is  in principle suited to study an orientational
transition like an instability due to a nematic phase \cite{letzallen99}.

\begin{figure}
\unitlength1cm
\epsfxsize=9.3cm
\begin{picture}(7,12)
\put(-0.6,-0.3){{\epsffile{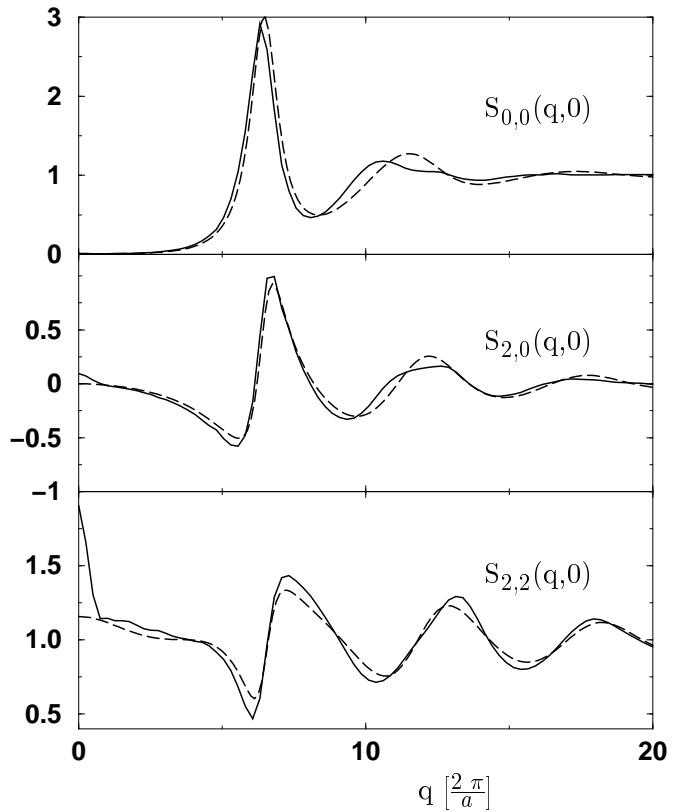}}}
\end{picture}
\caption{
Static structure factor for the A-B Lennard-Jones molecule (solid
lines) at the
glass transition T$_c$=0.475 obtained from
ref. \protect\cite{kaemmerer97}. 
The components of the static structure factor for  the hard ellipsoid of
revolution with an
aspect ratio of $X_0=1.5$ and with, at the glass transition, a packing
fraction of $\phi$= 0.575 are plotted with dashed lines.
From up to down the l=l'=0, m=0,
the l'=m=0, l=2 and the l=l'=2, m=0 components are shown.
}
\label{fig:1}
\end{figure}

In fig. \ref{fig:1} we have plotted with solid lines
the  three static correlators with $l=l'=0$; $l=0,l'=2$ and $l=l'=2$
for the Lennard--Jones system right at the ideal glass transition ($T_c
=0.475$ in L-J units and a density of $\rho_c = 0.752$)
. The first one ( $l=l'=0$) is the
center of mass component. The second component with $l=0$ and $l'=2$ couples
the two correlators for $l=l'=0$ and $l=l'=2$, whereas the third one, the
"quadrupolar" correlator gives information on the orientational order of
the system. In fig. \ref{fig:1} we have further plotted with dashed lines the
same components 
for the system of hard ellipsoids at the critical density (packing fraction
$\phi_c = 0.575$ or density $\rho_c = \frac{6 \phi}{\pi X_0} = 0.732$) of
the glass transition.
Note the remarkable similarity with the exception of the $q
\longrightarrow 0$ behaviour where the simulation result has some
shortcomings.\\[0.2cm]
These static correlators were used to solve the equations
(\ref{eq:mctanf})-(\ref{eq:mctend}) which are truncated at  $l_{max}=2$.
As already mentioned above, the A--B molecules do not possess head--tail symmetry.

\begin{figure}
\unitlength1cm
\epsfxsize=9.3cm
\begin{picture}(7,12)
\put(-0.6,-0.3){{\epsffile{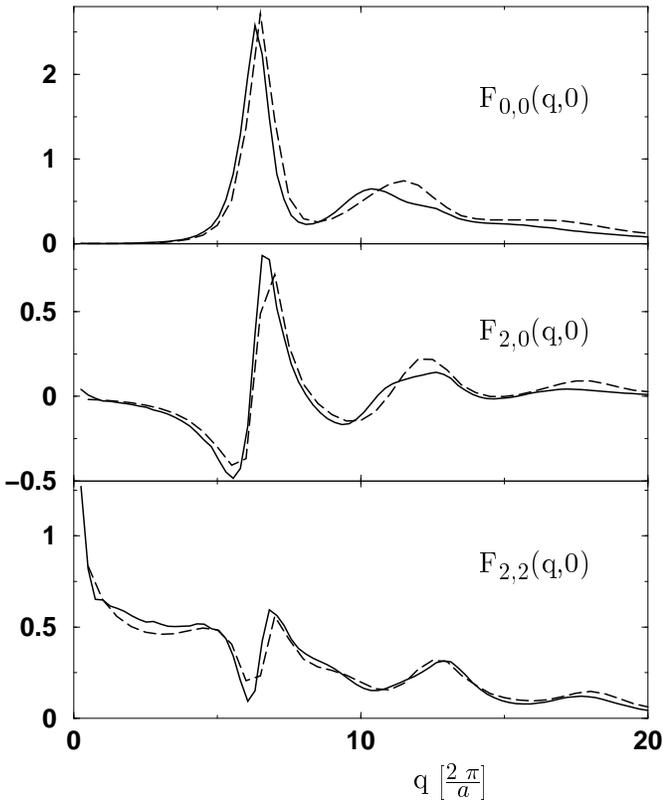}}}
\end{picture}
\caption{
Non-ergodicity parameter as they arise from solving the mode coupling
equations in the limit of $t \longrightarrow \infty$ for the A-B
Lennard-Jones molecule (solid lines). The non-ergodicity
parameters for the hard ellipsoids 
with an aspect ratio of $X_0=1.5$ and, at the glass transition,
a packing fraction $\phi$= 0.575
are plotted with dashed lines.
From up to down the l=l'=0, m=0,
the l'=m=0, l=2 and the l=l'=2, m=0 components are shown.
}
\label{fig:2}
\end{figure}

Therefore the static correlators with $l$ and/or $l'$ odd do not
vanish, in contrast to those for the ellipsoids. Accordingly these correlators
were 
also taken into account when solving the MCT--equations for the A--B 
molecules. The resulting critical non-ergodicity parameters $F_{ll'}(q,m)$
are plotted in fig. \ref{fig:2} for the  A-B LJ molecule with solid lines
for the ellipsoids with dashed lines. It is clearly visible that the
dominating contribution for the ergodicity breaking results from
the first maximum (at $q \approx 6.5 \left [ \frac{2 \pi}{a} \right
]$) of 
the structure factor of the center of mass
component ($l=l'=0$). This is the manifestation of the cage effect.  

\section{Conclusion}
In conclusion, we have demonstrated that a system of hard ellipsoids of
revolution is able to describe the static structure and the critical
non-ergodicity parameters (for $l$ and $l'$ even) of a system of super-cooled diatomic molecules
rather well, provided the aspect ratio is chosen properly. 
This result is not so obvious, since the static input into the
MCT--equations for the diatomic molecules also includes the
correlators with $l$ and/or $l'$ odd. Those correlators either equal
one or zero in case of the head--tail symmetric ellipsoids. 
For atoms with aspect ratios $X_0 \approx 1.5$ the cage effect
determines the glass transition. This relation however will
change close to a nematic instability.
On one particular example we have shown that the 
detailed relation between orientational and center of mass components
and the detailed shape and even the exact symmetry
of the two--particle potential seems only to have minor
influence on the ergodicity breaking. 
Of course, it would be interesting to investigate how far more complex
molecular systems can be modeled by hard ellipsoids. The full phase
diagram for the glass transition of hard 
ellipsoids will be presented elsewhere \cite{letzschil99}.

\acknowledgements

We acknowledge financial support from 
the Deutsche Forschungsgemeinschaft through
SFB 262.

\end{document}